\documentclass[preprint,preprintnumbers, prd, floatfix, superscriptaddress,nofootinbib] {revtex4-1}
\usepackage{epsfig}
\usepackage{subfigure}
\usepackage{dcolumn}% Align table columns on decimal point
\usepackage{bm}% bold math
\usepackage[usenames ,dvipsnames]{xcolor}
\usepackage{slashed}
\usepackage{graphicx,color}

\begin{document}
\title{Two-body charmed baryon decays involving vector meson
with $SU(3)$ flavor symmetry
}
\author{Y.K. Hsiao}
\email{yukuohsiao@gmail.com}
\affiliation{School of Physics and Information Engineering, Shanxi Normal University, Linfen 041004, China}

\author{Yu Yao}
\email{yuyao@cqupt.edu.cn}
\affiliation{Chongqing University of Posts \& Telecommunications, Chongqing, 400065, China}

\author{H.J. Zhao}
\affiliation{School of Physics and Information Engineering, Shanxi Normal University, Linfen 041004, China}

\begin{abstract}
We study the two-body anti-triplet charmed baryon decays
of ${\bf B}_c\to {\bf B}_n V$, with ${\bf B}_c=(\Xi_c^{0},-\Xi_c^{+},\Lambda_c^+)$
and ${\bf B}_n(V)$ the baryon (vector meson) states.
Based on the $SU(3)$ flavor symmetry, we predict that 
${\cal B}(\Lambda^{+}_{c}\to \Sigma^{+}\rho^{0},\Lambda^0 \rho^+)
=(0.61\pm 0.46,0.74\pm 0.34)\%$, 
in agreement with the experimental upper bounds of $(1.7,6)\%$,
respectively. 
We also find 
${\cal B}(\Lambda^+_c \to 
\Xi^0 K^{*+},\Sigma^0 K^{*+},\Lambda^0 K^{*+})
=(8.7 \pm 2.7,1.2\pm 0.3,2.0\pm 0.5)\times 10^{-3}$
to be compatible with the pseudoscalar counterparts.
For the doubly Cabibbo-suppressed decay $\Xi^+_c \to p\phi$,
measured for the first time,
we predict its branching ratio to be $(1.5\pm 0.7)\times 10^{-4}$,
together with 
${\cal B}(\Xi^+_c \to p \bar K^{*0},\Sigma^+ \phi)
=(7.8 \pm 2.2,1.9\pm0.9)\times 10^{-3}$.
The ${\bf B}_c\to{\bf B}_n V$ decays with 
${\cal B}\simeq {\cal O}(10^{-4}-10^{-3})$
are accessible to the BESIII, BELLEII and LHCb experiments.

\end{abstract}

\maketitle
\section{Introduction}
The two-body ${\bf B}_c\to{\bf B}_n V$ decays have not been abundantly
measured as the ${\bf B}_c\to{\bf B}_n M$ counterparts,
where ${\bf B}_c=(\Xi_c^{0},-\Xi_c^{+},\Lambda_c^+)$ 
are the anti-triplet charmed baryon states,
together with ${\bf B}_n$ and $V(M)$
the baryon and vector (pseudo-scalar) meson states, respectively.
For example, all Cabibbo-favored (CF) $\Lambda_c^+\to {\bf B}_n M$ decays
have been measured~\cite{pdg}, including
the recent BESIII observation for $\Lambda_c^+\to \Sigma^+\eta'$~\cite{Ablikim:2018czr},
whereas for the CF vector modes
only $\Lambda_c^+\to p \bar K^{*0},\Sigma^+\omega,\Sigma^+\phi$
have absolute branching fractions~\cite{pdg}.
In addition,
the first absolute branching ratio for the $\Xi_c^0$ decays
is $\Xi_c^0\to \Xi^-\pi^+$~\cite{Li:2018qak}, instead of any $\Xi_c^0\to {\bf B}_n V$ decays.

Nevertheless, the ${\bf B}_c\to{\bf B}_n V$ decays
are not less important than the ${\bf B}_c\to{\bf B}_n M$ counterparts.
First, the participations of BESIII, BELLEII and LHCb Collaborations 
are expected to 
make more accurate measurements for ${\bf B}_c\to{\bf B}_n V$, 
such as $\Lambda^{+}_{c}\to \Sigma^{+}\rho^{0},\Lambda^0 \rho^+$,
presented as
${\cal B}(\Lambda_c^+\to \Sigma^{+}\rho^{0},\Lambda^0 \rho^+)<(1.7,6)\%$
due to the previous measurements~\cite{pdg}.
Second, in the three-body ${\bf B}_c\to {\bf B}_nMM'$ decays,
the $MM'$ meson pair is assumed to be mainly in the S-wave state~\cite{Geng:2018upx}.
However,
the resonant ${\bf B}_c\to{\bf B}_n V, V\to MM'$ decay
causes $MM'$ to be in the P-wave state, of which
the contribution to the total ${\cal B}({\bf B}_c\to {\bf B}_nMM')$ needs clarification.
Note that (S,P) denote $L=(0,1)$ as the quantum numbers
for the orbital angular momentum between $M$ and $M'$.
Third, the three-body $\Xi_c^+$ decays
can be measured as the ratios of
${\cal B}(\Xi_c^+\to{\bf B}_n V)/{\cal B}(\Xi_c^+\to {\bf B}_nMM')$.
Particularly, the doubly Cabibbo-suppressed
$\Xi_c^+\to p\phi$ decay is observed for the first time, with
${\cal B}(\Xi_c^+\to p\phi)/{\cal B}(\Xi_c^+\to pK^-\pi^+)
=(19.8\pm 0.7\pm 0.9\pm 0.2)\times 10^{-3}$~\cite{Aaij:2019kss}.
The information of ${\cal B}(\Xi_c^+\to{\bf B}_n V)$ is hence helpful
to determine ${\cal B}(\Xi_c^+\to {\bf B}_nMM')$.

Since the study of ${\bf B}_c\to{\bf B}_n V$ is necessary,
it is important to provide a corresponding theoretical approach.
The factorization approach for 
the heavy hadron decays~\cite{ali,Geng:2006jt,Hsiao:2014mua}
seems applicable to ${\bf B}_c\to{\bf B}_n V$.
Nonetheless, it has been shown that,
besides the factorizable effects, there exist significant
non-factorizable contributions in ${\bf B}_c\to{\bf B}_nM$~\cite{Zhao:2018mov}, 
such that the factorization approach fails to explain the data.
In contrast, with both
factorizable and non-factorizable effects~\cite{He:2000ys,Fu:2003fy,Hsiao:2015iiu,
He:2015fwa,He:2015fsa,Savage:1989qr,Savage:1991wu,h_term,He:2018joe},
the $SU(3)$ flavor symmetry ($SU(3)_f$) approach
can accommodate the measurements for 
${\bf B}_c\to{\bf B}_nM$~\cite{Lu:2016ogy,Wang:2017azm,Wang:2017gxe,
Geng:2017esc,Geng:2018bow,Geng:2018plk,Geng:2018rse}, 
such as
the purely non-factorizable $\Lambda_c^+\to\Xi^0 K^+$ decay~\cite{Ablikim:2018bir}.
In addition, the predicted values of
${\cal B}(\Lambda_c^+\to \Sigma^+\eta')$ and
${\cal B}(\Xi_c^0 \to \Xi^-\pi^+)$ are in agreement with
the recent observations~\cite{Ablikim:2018czr,Li:2018qak,Geng:2017esc,Geng:2018plk}.
Therefore, we propose to extend the $SU(3)_f$ symmetry
to ${\bf B}_c\to{\bf B}_nV$,
while the existing observations have been sufficient 
for the numerical analysis.
In this report, we will extract the $SU(3)_f$ amplitudes, and
predict the to-be-measured ${\bf B}_c\to{\bf B}_nV$ branching fractions.

\section{Formalism}
To obtain the amplitudes
for the two-body ${\bf B}_c\to {\bf B}_n V$ decays,
where ${\bf B}_{c(n)}$ is the singly charmed (charmless) baryon state
and $V$ the vector meson,
we present the relevant effective Hamiltonian $({\cal H}_{eff})$
for the tree-level $c$ quark decays, given by~\cite{Buras:1998raa}
\begin{eqnarray}\label{Heff}
{\cal H}_{eff}&=&\sum_{i=+,-}\frac{G_F}{\sqrt 2}c_i
\left(V_{cs}V_{ud}O_i+V_{cq}V_{uq} O_i^q+V_{cd}V_{us}O'_i\right),
\end{eqnarray}
with $q=d$ or $s$, where $G_F$ is the Fermi constant, 
$V_{ij}$ are the CKM matrix elements, and
$c_{\pm}$ the scale-dependent Wilson coefficients. In Eq.~(\ref{Heff}),
$O_\pm^{(q,\prime)}$ are the four-quark operators:
\begin{eqnarray}\label{O12}
&&
O_\pm={1\over 2}\left[(\bar u d)(\bar s c)\pm (\bar s d)(\bar u c)\right]\,,\;\nonumber\\
&&
O_\pm^q={1\over 2}\left[(\bar u q)(\bar q c)\pm (\bar q q)(\bar u c)\right]\,,\;\nonumber\\
&&
O'_\pm={1\over 2}\left[(\bar u s)(\bar d c)\pm (\bar d s)(\bar u c)\right]\,,
\end{eqnarray}
with $(\bar q_1 q_2)\equiv \bar q_1\gamma_\mu(1-\gamma_5)q_2$.
By neglecting the Lorentz indices,
the operator of $(\bar q_1q_2)(\bar q_3 c)$
transforms as $(\bar q^i q_k \bar q^j)c$ under the $SU(3)_f$ symmetry, where
$q_i=(u,d,s)$ represent the triplet of $3$.
The operator can be decomposed as irreducible forms, 
which is accordance with
$(\bar 3\times 3\times \bar 3)c=(\bar 3+\bar 3'+6+\overline{15})c$.
One hence has~\cite{Savage:1989qr,Savage:1991wu}
\begin{eqnarray}\label{irreOi}
&&O_{-(+)}\sim  {\cal O}_{6(\overline{15})}=
\frac{1}{2}(\bar u d\bar s\mp\bar s d\bar u)c\,,\nonumber\\
&&O_{-(+)}^q \sim{\cal O}_{6(\overline{15})}^q=
{1\over 2}(\bar u q\bar q\mp \bar q q\bar u)c\,,\nonumber\\
&&O'_{-(+)}\sim  {\cal O'}_{6(\overline{15})}=
{1\over 2}(\bar u s\bar d\mp \bar d s\bar u)c\,,
\end{eqnarray}
with the subscripts $(6,\overline{15})$ denoting
the two irreducible $SU(3)_f$ operators.
By substituting   
${\cal O}_{6(\overline{15})}^{(q,\prime)}$ for $O_{-(+)}^{(q,\prime)}$,
the effective Hamiltonian in Eq.~(\ref{Heff}) becomes
\begin{eqnarray}\label{Heff2}
{\cal H}_{eff}&=&\frac{G_F}{\sqrt 2}\left[c_- { \epsilon^{ijl} \over 2}H(6)_{lk}+c_+H(\overline{15})_k^{ij}\right]\,,
\end{eqnarray}
where the tensor notations of
$1/2\epsilon_{ijl}H(6)^{lk}$ and $H(\overline{15})_{ij}^k$
contain ${\cal O}_{6}^{(q,\prime)}$ and ${\cal O}_{\overline{15}}^{(q,\prime)}$,
respectively. In terms of 
$(V_{cs}V_{ud},V_{cd}V_{ud},V_{cs}V_{us},V_{cd}V_{us})=(1,-s_c,s_c,-s_c^2)$
with $s_c\equiv \sin\theta_c$, 
where $\theta_c$ represents  the well-known Cabbibo angle,
we have $H_{22}(6)=2$,
$H^{23,32}(6)=-2s_c$, $H^2_{12,21}(\overline{15})=-H^3_{13,31}(\overline{15})=s_c$,
$H^{33}(6)=2s_c^2$, and $H^3_{12,21}(\overline{15})=-s_c^2$
as the non-zero entries~\cite{Savage:1989qr}.
Note that $n=0,1$ and 2 in $s_c^n$ correspond to
the Cabibbo-flavored (CF), singly Cabibbo-suppressed (SCS), and
doubly Cabibbo-suppressed (DCS) decays, respectively.
We also need ${\bf B}_c$ and ${\bf B}_n$ ($V$) to be
in the irreducible representation of the $SU(3)_f$ symmetry, 
given by
\begin{eqnarray}\label{b_octet}
({\bf B}_{c})_i&=&(\Xi_c^0,-\Xi_c^+,\Lambda_c^+)\,,\nonumber\\
({\bf B}_n)^i_j&=&\left(\begin{array}{ccc}
\frac{1}{\sqrt{6}}\Lambda^0+\frac{1}{\sqrt{2}}\Sigma^0 & \Sigma^+ & p\\
 \Sigma^- &\frac{1}{\sqrt{6}}\Lambda^0 -\frac{1}{\sqrt{2}}\Sigma^0  & n\\
 \Xi^- & \Xi^0 &-\sqrt{\frac{2}{3}}\Lambda^0
\end{array}\right)\,,\nonumber\\
(V)^i_j&=&\left(\begin{array}{ccc}
\frac{1}{\sqrt{2}}(\rho^0+ \omega) & \rho^- & K^{*-}\\
 \rho^+ & \frac{-1}{\sqrt{2}}(\rho^0-\omega) & \bar K^{*0}\\
 K^{*+} & K^{*0}& \phi
\end{array}\right)\,.
\end{eqnarray}
Subsequently, 
${\cal H}_{eff}$ in Eq.~(\ref{Heff2}) is enabled to be connected
to the initial and final states in Eq.~(\ref{b_octet}),
such that we derive
the amplitudes of ${\bf B}_c\to {\bf B}_n V$ as
\begin{eqnarray}
&&{\cal A}({\bf B}_c\to {\bf B}_n V)
=\langle {\bf B}_n V|{\cal H}_{eff}|{\bf B}_c\rangle
=\frac{G_F}{\sqrt 2}T({\bf B}_{c}\to {\bf B}_nV)\,,
\end{eqnarray}
instead of introducing the details of the QCD calculations
for the hadronization. Explicitly, the $T$ amplitudes ($T$-amps) 
are given by~\cite{Savage:1989qr,Savage:1991wu}
\begin{eqnarray}\label{Tamp}
T({\bf B}_c\to {\bf B}_n V)&=&T({\cal O}_6)+T({\cal O}_{\overline{15}})\,,\nonumber\\
T({\cal O}_6)&=&
\bar a_1 H^{ij}(6)T_{ik}({\bf B}_n)^k_l (V)^l_j+
\bar a_2 H^{ij}(6)T_{ik}(V)^k_l ({\bf B}_n)^l_j\nonumber\\
&+&
\bar a_3 H^{ij}(6)({\bf B}_n)^k_i (V)^l_j T_{kl}+
\bar h H^{ij}(6)T_{ik}({\bf B}_n)^k_j (V)^l_l\,,\nonumber\\
T({\cal O}_{\overline{15}})&=&
\bar a_4H^{i}_{jk}(\overline{15}) (V)^j_l ({\bf B}_n)^k_i({\bf B}_{c})^l
+\bar a_5 H(\overline{15})^{i}_{jk} ({\bf B}_{c})^j({\bf B}_n)^k_l (V)^l_i\nonumber\\
&+&
\bar a_6H(\overline{15})^i_{jk}({\bf B}_n)^j_l (V)^k_i  ({\bf B}_{c})^l
+\bar a_7 H(\overline{15})^i_{jk}({\bf B}_{c})^j (V)^k_l({\bf B}_n)^l_i \nonumber\\
&+&
\bar h' H^{i}_{jk}(\overline{15})({\bf B}_n)^k_i (V)^l_l ({\bf B}_{c})^j\,,
\end{eqnarray}
where $T_{ij} \equiv ({\bf B}_c)_k\epsilon^{ijk}$, and
$(c_-,c_+)$ have been absorbed into the $SU(3)$ parameters
$(\bar a_i,\bar h^{(\prime)})$.
%
%=======================================================
\begin{table}[t!]
\caption{The $T$-amps for the ${\bf B}_{c}\to {\bf B}_n V$ decays, where
CF denotes the Cabibbo-favored processes, while
SCS (DCS)  the singly (doubly) Cabibbo-suppressed ones.}\label{tab1}
{%\tiny
\scriptsize
%\footnotesize
%=====================
%==============================
\begin{tabular}{|c|l|}
\hline
$\Xi_c^0$&$\;\;\;\;\;\;\;\;$CF $T$-amp
\\
\hline
$\Sigma^{+} K^{*-} $
& $ 2(\bar a_{2}+\frac{\bar a_{4} + \bar a_{7}}{2})$
\\
$\Sigma^{0}\bar{K}^{*0}$
&$-\sqrt{2}(\bar a_{2}+\bar a_{3}$
$-\frac{\bar a_{6}-\bar a_{7}}{2})$
\\
$\Xi^{0} \rho^{0} $
& $ -\sqrt{2}(\bar a_{1}-\bar a_{3}$
$-\frac{\bar a_{4}-\bar a_{5}}{2})$
\\
$ \Xi^{0} \omega $ & $ \sqrt{2}(\bar a_1- \bar a_3 +2\bar h $
$+ \frac{\bar a_4+\bar a_5+2 \bar h'}{2} )$\\

$ \Xi^{0} \phi $
&$\bar a_2 +\bar h + \frac{\bar a_7 + \bar h'}{2}  $ \\
$\Xi^{-} \rho^{+} $
& $ 2(\bar a_{1}+\frac{\bar a_{5} + \bar a_{6}}{2})$
\\
$\Lambda^{0} \bar{K}^{*0} $
&
$-\sqrt{\frac{2}{3}}(2\bar a_1-\bar a_2-\bar a_3$
\\
&
$+\frac{2\bar a_5-\bar a_6-\bar a_7}{2})$
\\[40mm]

\hline\hline
$\Xi_c^+$&$\;\;\;\;\;\;\;\;$CF $T$-amp
\\\hline
$\Sigma^{+} \bar{K}^{*0} $
&$ -2(\bar a_{3}-\frac{\bar a_{4} + \bar a_{6}}{2})$
\\
$\Xi^{0} \rho^{+} $
& $2(\bar a_{3}+\frac{\bar a_{4} + \bar a_{6}}{2})$
\\[45mm]
\hline\hline
$\Lambda_c^+$&$\;\;\;\;\;\;\;\;$CF $T$-amp
\\\hline
$\Sigma^{0} \rho^{+} $
&$-\sqrt{2}(\bar a_1-\bar a_2-\bar a_3$
$-\frac{\bar a_5-\bar a_7}{2})$
\\
$\Sigma^{+} \rho^{0} $
& $\sqrt{2}(\bar a_{1}-\bar a_{2}-\bar a_{3}$
$-\frac{\bar a_{5}-\bar a_{7}}{2})$
\\
$ \Sigma^{+} \omega $ & $ \sqrt{2}(-\bar a_1-\bar a_2+\bar a_3-2\bar h$ \\
&$+\frac{\bar a_5+\bar a_7+2\bar h'}{2}) $ \\

$ \Sigma^{+} \phi $
&$ \bar a_4-2\bar h+\bar h'  $
\\
$\Xi^{0} K^{*+} $
& $-2(\bar a_{2}-\frac{\bar a_{4} + \bar a_{7}}{2})$
\\
$p \bar{K}^{*0} $
& $ -2(\bar a_{1}-\frac{\bar a_{5} + \bar a_{6}}{2})$
\\
$\Lambda^{0} \rho^{+} $
&
$-\sqrt{\frac{2}{3}}(\bar a_1+\bar a_2+\bar a_3$
\\
&$-\frac{\bar a_5-2\bar a_6+\bar a_7}{2})$
\\[5mm]
\hline

\end{tabular}
%============================
\begin{tabular}{|c|l|}
\hline
$\Xi_{c}^{0}$ &$\;\;\;\;\;\;\;\;\;$SCS $T$-amp
\\\hline

$\Sigma^{+} \rho^{-} $
& $-2(\bar a_{2}+\frac{\bar a_{4} + \bar a_{7}}{2})s_c$
\\
$\Sigma^{-} \rho^{+} $
& $-2(\bar a_{1}+\frac{\bar a_{5} + \bar a_{6}}{2})s_c$
\\
$\Sigma^{0} \rho^{0} $
&$-(\bar a_{2}+\bar a_{3}$
$-\frac{\bar a_{4}-\bar a_{5}+\bar a_{6}-\bar a_{7}}{2})s_c$
\\
$ \Sigma^{0} \omega $
& $[-(\bar a_1+\bar a_2+2\bar h$\\
&
$+\frac{\bar a_4+\bar a_5-\bar a_6+\bar a_7+2\bar h'}{2})
]s_c$
\\
$ \Sigma^{0} \phi $
& $[\sqrt{2}(\bar a_3-\bar h-\frac{\bar a_6+\bar h'}{2})]s_c$
\\
$\Xi^{-} K^{*+} $
& $ 2(\bar a_{1}+\frac{\bar a_{5} + \bar a_{6}}{2})s_c$
\\
$p K^{*-}$
&$2(\bar a_{2}+\frac{\bar a_{4} + \bar a_{7}}{2})s_c$
\\
$\Xi^{0} K^{*0} $
& $2(\bar a_{1}-\bar a_2-\bar a_{3}$
$+\frac{\bar a_{5}-\bar a_{7}}{2})s_c$
\\
$n \bar K^{*0} $
& $-2(\bar a_{1}-\bar a_{2}-\bar a_{3}+\frac{\bar a_{5}-\bar a_{7}}{2})s_c$
\\
$\Lambda^{0} \rho^{0} $
&
$\sqrt{\frac{1}{3}}(\bar a_1+\bar a_2-2\bar a_3$
\\
&$+\frac{\bar a_4-\bar a_5-\bar a_6-\bar a_7}{2})s_c$
\\
$ \Lambda^{0} \omega $
& $ \frac{\sqrt{3}}{3}(\bar a_1+\bar a_2-2\bar a_3+6\bar h $\\
&$+\frac{3\bar a_4+\bar a_5+\bar a_6+\bar a_7+6\bar h'}{2}) s_c$
\\
$ \Lambda^{0} \phi $
&$\frac{\sqrt{6}}{2}(2\bar a_1+2\bar a_2-\bar a_3+3\bar h $\\
&$+\frac{2\bar a_5-\bar a_6+2\bar a_7+3\bar h'}{2})s_c $
\\

\hline\hline
$\Xi_{c}^{+}$ &$\;\;\;\;\;\;\;\;\;$SCS $T$-amp
\\\hline
$\Sigma^{0} \rho^{+} $
&$\sqrt 2(\bar a_{1}-\bar a_{2}$
%&$\sqrt 2(a_{1}-a_{2}+\frac{a_{4}-a_{5}+a_{6}+a_{7}}{2})s_c$
\\
&$+\frac{\bar a_{4}-\bar a_{5}+\bar a_{6}+\bar a_{7}}{2})s_c$
\\
$\Sigma^{+} \rho^{0} $
&$-\sqrt 2(\bar a_{1}-\bar a_{2}$
\\
&$-\frac{\bar a_{4}+\bar a_{5}+\bar a_{6}-\bar a_{7}}{2})s_c$
\\
$ \Sigma^{+} \omega $
& $\sqrt{2}(\bar a_1+\bar a_2+2\bar h$ \\
&$-\frac{\bar a_4+\bar a_5+\bar a_6+\bar a_7-2\bar h'}{2})
s_c$
\\
$ \Sigma^{+} \phi $
&$-2(\bar a_3-\bar h-\frac{\bar a_6-\bar h'}{2})s_c$
\\
$\Xi^{0} K^{*+} $
& $2(\bar a_2+\bar a_{3}+\frac{\bar a_{6} - \bar a_{7}}{2})s_c$
\\
$p \bar K^{*0} $
& $2(\bar a_1-\bar a_{3}+\frac{\bar a_{4} - \bar a_{5}}{2})s_c$
\\
$\Lambda^0\rho^+$
& $\sqrt{\frac{2}{3}}(\bar a_1+\bar a_2-2\bar a_3$
\\
&$-\frac{3\bar a_4+\bar a_5+\bar a_6+\bar a_7}{2})s_c$
\\

\hline\hline
$\Lambda_{c}^{+}$ &$\;\;\;\;\;\;\;\;\;$SCS $T$-amp
\\\hline

$\Sigma^{+} K^{*0} $
& $-2(\bar a_{1}-\bar a_{3}-\frac{\bar a_{4}-\bar a_{5}}{2})s_c$
\\
$\Sigma^{0} K^{*+} $
&$-\sqrt{2}(\bar a_1-\bar a_3-\frac{\bar a_4+\bar a_5}{2})s_c$
\\
$p \rho^{0} $
& $ -\sqrt 2(\bar a_{2}+\bar a_3-\frac{\bar a_{6} - \bar a_{7}}{2})s_c$
\\
$ p \omega $
&$\sqrt{2}(\bar a_2-\bar a_3+2\bar h$\\
&$+\frac{\bar a_6-\bar a_7-2\bar h'}{2})s_c $
\\
$ p \phi $
&$ -2(-\bar a_1-\bar h$\\
&$+\frac{\bar a_4+\bar a_5+\bar a_6+\bar h'}{2})s_c$
\\
$n\rho^+$
&$-2(\bar a_{2}+\bar a_3-\frac{\bar a_{4} + \bar a_{7}}{2})s_c$
\\
$\Lambda^{0} K^{*+} $
&
$-\sqrt{\frac{2}{3}}(\bar a_1-2\bar a_2+\bar a_3$
\\
&$-\frac{3\bar a_4-\bar a_5+2\bar a_6+2\bar a_7}{2})s_c$
\\
\hline
\end{tabular}
%=====================
\begin{tabular}{|c|l|}
\hline
$\Xi_c^0$&$\;\;\;\;\;\;\;\;$DCS $T$-amp
\\\hline
%\hline
$p\rho^-$&
$-2(\bar a_{2}+\frac{\bar a_{4} + \bar a_{7}}{2})s_c^2$
\\
$\Sigma^{-} K^{*+} $
& $-2(\bar a_{1}+\frac{\bar a_{5} + \bar a_{6}}{2})s_c^2$
\\
$\Sigma^{0}{K}^{*0}$
&$ \sqrt 2(\bar a_{1}+\frac{\bar a_{5} - \bar a_{6}}{2})s_c^2$
\\
$n \rho^{0} $
&$\sqrt 2(\bar a_{2}-\frac{\bar a_{4} - \bar a_{7}}{2})s_c^2$
\\
$ n \omega $
&$-\sqrt{2}( \bar a_2-2\bar h + \frac{\bar a_4 -\bar a_7 -2\bar h'}{2} )
s_c^2$ \\
$ n \phi $

&$-2( \bar a_1 -\bar a_3 +\bar h +\frac{\bar a_5+\bar h'}{2})s_c^2$ \\
$\Lambda^{0} {K}^{*0} $
&
$-\sqrt{\frac{2}{3}}(\bar a_1-2\bar a_2-2\bar a_3$\\
&
$+\frac{\bar a_5+\bar a_6-2\bar a_7}{2})s_c^2$
\\[40mm]
\hline\hline
$\Xi_c^+$&$\;\;\;\;\;\;\;\;$DCS $T$-amp
\\\hline
$\Sigma^{0} {K}^{*+} $
&$ \sqrt 2(\bar a_{1}-\frac{\bar a_{5} - \bar a_{6}}{2})s_c^2$
\\
$\Sigma^{+} {K}^{*0} $
&$ 2(\bar a_{1}-\frac{\bar a_{5} + \bar a_{6}}{2})s_c^2$
\\
$p \rho^0 $
&$\sqrt 2(\bar a_{2}+\frac{\bar a_{4} - \bar a_{7}}{2})s_c^2$
\\
$ p \omega $
%& $[\sqrt{2}(-a_2 + 2h $ \\
& $[\sqrt{2}(-\bar a_2 + 2\bar h+\frac{\bar a_4+\bar a_7 + 2\bar h'}{2})] s_c^2 $ \\
%&$+\frac{a_4+a_7 + 2h'}{2})] s_c^2$
%\\
$ p \phi $
&$ -2(\bar a_1 - \bar a_3 +h- \frac{\bar a_5 +\bar h'}{2})s_c^2$  \\
%&$ - \frac{a_5 +h'}{2})s_c$
%\\
$n \rho^{+} $
& $2(\bar a_{2}-\frac{\bar a_{4} + \bar a_{7}}{2})s_c^2$
\\
$\Lambda^0 K^{*+}$
& $\sqrt{\frac{2}{3}}(\bar a_1-2\bar a_2-2\bar a_3$
\\
&$-\frac{\bar a_5+\bar a_6-2\bar a_7}{2})s_c^2$
\\[15mm]
\hline\hline
$\Lambda_c^+$&$\;\;\;\;\;\;\;\;$DCS $T$-amp
\\\hline
$p {K}^{*0} $
& $ 2(\bar a_{3}-\frac{\bar a_{4} + \bar a_{6}}{2})s_c^2$
\\
$nK^{*+}$&
$- 2(\bar a_{3}+\frac{\bar a_{4} + \bar a_{6}}{2})s_c^2$
\\[40mm]
\hline
\end{tabular}
}
\end{table}
%end of the table==========================================
%
%
With the full expansion of $T$-amps in Table~\ref{tab1},
the two-body ${\bf B}_c\to {\bf B}_nV$ decays
are presented with the $SU(3)_f$ parameters.
Since $\omega=(u\bar u+d\bar d)/\sqrt 2$ and
$\phi=s\bar s$ actually mix with
$\omega_1=(u\bar u+d\bar d+s\bar s)/\sqrt 3$ and
$\omega_8=(u\bar u+d\bar d-2s\bar s)/\sqrt 6$,
the $(\bar h,\bar h')$ terms that are related to
$(V)^l_l =\sqrt 2 \omega+\phi=\sqrt 3 \omega_1$
can contribute to the decays with $(\omega,\phi)$ only. 
In terms of the equation for the two-body decays,
given by~\cite{pdg}
\begin{eqnarray}\label{p_space}
&&{\cal B}({\bf B}_c\to {\bf B}_n V)=
\frac{|\vec{p}_{cm}|\tau_{\bf{B}_c}}{8\pi m_{{\bf B}_c}^2 }|{\cal A}({\bf B}_c\to {\bf B}_n V)|^2\,,\nonumber\\
&&|\vec{p}_{cm}|=\frac{
\sqrt{[(m_{{\bf B}_c}^2-(m_{{\bf B}_n}+m_V)^2]
[(m_{{\bf B}_c}^2-(m_{{\bf B}_n}-m_V)^2]}}{2 m_{{\bf B}_c}}\,,
\end{eqnarray}
we can compute the branching ratio with the $SU(3)_f$ amplitudes,
where $\tau_{\bf{B}_c}$ denotes the $\bf{B}_c$ lifetime.
The $SU(3)_f$ amplitudes are accounted to be
9 complex numbers, leading to 17 independent parameters 
to be extracted, whereas there exist 10 data points 
for the numerical analysis.
To have a practical fit, we follow 
Refs.~\cite{Lu:2016ogy,Geng:2017esc,Geng:2018plk,Geng:2018rse} 
to reduce the parameters.
In ${\cal H}_{eff}\propto c_-H(6)+c_+H(\overline{15})$,
since the QCD calculation at the scale $\mu=1$ GeV
leads to $(c_+,c_-)=(0.76,1.78)$ in the naive
dimensional regularization (NDR) scheme~\cite{Li:2012cfa,Fajfer:2002gp},
the ratio of $(c_-/c_+)^2\simeq 0.17$ indicates the suppression of $H(\overline{15})$.
We hence ignore $(\bar a_{4,5, ...,7},h')$.
On the other hand, $(\bar a_{1,2,3},h)$ from $H(6)$ are kept for the fit,
represented as
\begin{eqnarray}\label{7p}
\bar a_1, \bar a_2e^{i\delta_{\bar a_2}},\bar a_3e^{i\delta_{\bar a_3}},
\bar h e^{i\delta_{\bar h}}\,,
\end{eqnarray}
with the phases $\delta_{\bar a_{2,3},\bar h}$, 
and $\bar a_1$ set to be relatively real.

\section{Numerical analysis}

For the numerical analysis,
we collect (the ratios of) the branching fractions
for the observed ${\bf B}_c\to{\bf B}V$ decays
in Table~\ref{data}, 
where ${\cal B}(\Xi^+_c \to p \bar K^{*0},\Sigma^+ \phi,\Sigma^+\bar K^{*0})$
are in fact measured to be relative to ${\cal B}(\Xi_c^+\to\Xi^-\pi^+\pi^+)$~\cite{pdg,Link:2003cd},
recombined as ${\cal R}_{1,2}(\Xi^+_c)$.
We obtain ${\cal B}( \Xi^0_c \to \Lambda^0 \phi)$ from
${\cal B}( \Xi^0_c \to \Lambda^0 \phi)/{\cal B}( \Xi^0_c \to \Xi^-\pi^+)$~\cite{pdg},
with the input of ${\cal B}( \Xi^0_c \to \Xi^- \pi^+)$
measured by BELLE~\cite{Li:2018qak}.
In addition,
the ratio of ${\cal R}(\Lambda_c^+)=
(\Lambda_c^+ \to \Sigma^+ \rho^0)/{\cal B}(\Lambda_c^+\to \Sigma^+ \omega)$
comes from the data events in Ref.~\cite{Kubota:1993pw}.
Besides, $s_c=0.22453\pm 0.00044$~\cite{pdg}
is the theoretical input for the CKM matrix elements.
By using the equation of~\cite{Zhao:2018mov}
\begin{eqnarray}\label{chi_eq}
\chi^2=
\sum_{i} \bigg(\frac{{\cal B}^i_{th}-{\cal B}^i_{ex}}{\sigma_{ex}^i}\bigg)^2+
\sum_{j}\bigg(\frac{{\cal R}^j_{th}-{\cal R}^j_{ex}}{\sigma_{ex}^j}\bigg)^2\,,
\end{eqnarray}
we are able to obtain the minimum $\chi^2$ value, such that
the $SU(3)_f$ parameters can be extracted with the best fit.
Note that ${\cal B}^i({\cal R}^j)$ represents (the ratio of) the branching fraction,
with the subscript $th$ ($ex$) denoting the theoretical (experimental) input,
while $\sigma_{ex}^{i(j)}$ stands for the experimental error.
As the inputs in Eq.~(\ref{chi_eq}),
${\cal B(R)}_{th}$ come from the $T$-amps in Table~\ref{tab1},
%with the ignoring of $(a_{4,5,...,7},h')$, 
while
${\cal B(R)}_{ex}$ and $\sigma_{ex}$ the data points in Table~\ref{data}.
%================================
\begin{table}[t!]
\caption{The (ratios of) branching fractions of the ${\bf B}_c\to {\bf B}_n V$ decays.
In column~2, the numbers are calculated with the extracted parameters,
in comparison with the initial experimental inputs in column~3.
}\label{data}
\begin{tabular}{|c|c|c|}
\hline
(Ratio of) Branching fraction
&This work&Data\\
\hline
\hline
$10^2{\cal B}(\Lambda_c^+ \to p \bar K^{*0})$
&$1.9 \pm 0.3$
&$1.94\pm 0.27$~\cite{pdg}\\
$10^2{\cal B}(\Lambda_c^+ \to \Sigma^+ \omega)$
&$1.6 \pm0.7$
&$1.69\pm 0.21$~\cite{pdg}\\
$10^3{\cal B}(\Lambda_c^+ \to \Sigma^+ \phi)$
&$3.9\pm0.6$
&$3.8\pm 0.6$~\cite{pdg}\\
${\cal R}(\Lambda^+_c )=
\frac{{\cal B}( \Lambda_c^+ \to \Sigma^+ \rho^0)}{{\cal B}(\Lambda_c^+ \to \Sigma^+ \omega)}$
&$0.4\pm 0.3$
&$0.3\pm 0.2$~\cite{Kubota:1993pw}\\
\hline
$10^3{\cal B}(\Lambda_c^+ \to \Sigma^+ K^{*0})$
&$2.3 \pm 0.6$
&$3.4\pm 1.0$~\cite{pdg}\\
%\hline\hline
$10^4{\cal B}(\Lambda_c^+ \to p \omega)$
&$11.4\pm5.4$
&$9.4\pm 3.9$~\cite{Aaij:2017nsd}\\
$10^4{\cal B}(\Lambda_c^+ \to p \phi)$
&$10.4\pm 2.1$
&$10.6\pm 1.4$~\cite{pdg}\\
%\hline
\hline
$10^4{\cal B}( \Xi^0_c \to \Lambda^0 \phi)$
&$8.4\pm3.9$
&$6.1\pm 2.2$~\cite{pdg,Li:2018qak}\\
${\cal R}_1(\Xi^+_c )=\frac{{\cal B}( \Xi^+_c \to p \bar K^{*0})}{{\cal B}(\Xi^+_c \to \Sigma^+\bar K^{*0})}$
%&$0.08\pm0.01$
%&$0.14\pm 0.05$~\cite{pdg}\\
&$(1.6\pm0.2)s_c^2$
&$(2.8\pm1.0)s_c^2$~\cite{pdg}\\
${\cal R}_2(\Xi^+_c )=\frac{{\cal B}( \Xi^+_c \to \Sigma^+ \phi)}{{\cal B}(\Xi^+_c \to \Sigma^+\bar K^{*0})}$
%&{\color{red}$0.019\pm0.004$}
&$(0.4\pm0.1)s_c^2$
%&$0.086\pm 0.063$~\cite{pdg}\\
&$(1.7\pm 1.2)s_c^2$~\cite{pdg,Link:2003cd}\\
\hline
\end{tabular}
\end{table}
%==================================
%
Subsequently, the global fit gives
\begin{eqnarray}\label{su3_fit}
&&(\bar a_1,\bar a_2,\bar a_3,\bar h)=(0.22\pm 0.02,0.23\pm 0.04,
0.39\pm 0.05,0.16\pm0.01)\,\text{GeV}^3\,,\nonumber\\
&&(\delta_{\bar a_2},\delta_{\bar a_3},\delta_{\bar h})=
(-85.5\pm13.0,78.4\pm 8.8,99.3\pm 7.7)^\circ\,,\nonumber\\
&&\chi^2/n.d.f=6.3/3=2.1\,,
\end{eqnarray}
where $n.d.f$ represents the number of the degree of freedom.
With the fit results in Eq.~(\ref{su3_fit}), we calculate
the branching ratios, ${\cal R}(\Lambda_c^+)$ and ${\cal R}_{1,2}(\Xi_c^+)$
to be compared to their data inputs in Table~\ref{data}. Moreover,
we predict the branching fractions 
for the ${\bf B}_c\to{\bf B}_n V$ decays, 
given in Table~\ref{tab_result}.

%\newpage
%================================
\begin{table}[t!]
\caption{The numerical results of
the ${\bf B}_{c}\to {\bf B}_n V$ decays,
with ${\cal B}_{{\bf B}_nV}\equiv {\cal B}({\bf B}_c\to {\bf B}_nV)$.
}\label{tab_result}
{
%\tiny
%\scriptsize
%\footnotesize
%==============================
\begin{tabular}{|c|c|}
\hline
$\Xi_c^0$&our results
\\
\hline
$10^3{\cal B}_{\Sigma^{+} K^{*-}}$
&$9.3 \pm 2.9$\\
$10^3{\cal B}_{\Sigma^{0} \bar{K}^{*0}}$
&$2.7 \pm 2.2$\\
$10^2{\cal B}_{\Xi^{0} \rho^{0}}$
&$1.4 \pm 0.4$\\
$10^3{\cal B}_{\Xi^{0} \omega}$
&$1.0^{+8.6}_{-1.0}$\\
$10^4{\cal B}_{\Xi^{0} \phi}$
&$1.5 ^{+7.1}_{-1.5}$\\
$10^3{\cal B}_{\Xi^{-} \rho^{+}}$
&$8.6 \pm 1.2$\\
$10^3{\cal B}_{\Lambda^{0} \bar{K}^{*0}}$
&$4.6 \pm 2.1$\\
&\\&\\&\\&\\
\hline\hline
$\Xi_c^+$&our results
\\\hline
$10^2{\cal B}_{\Sigma^{+} \bar{K}^{*0}}$
&$10.1\pm 2.9$ \\
$10^2{\cal B}_{\Xi^{0} \rho^{+}}$
&$9.9 \pm2.9 $\\
&\\&\\&\\&\\&\\
\hline\hline
$\Lambda_c^+$&our results
\\\hline
$10^3{\cal B}_{\Sigma^{0} \rho^{+}}$
&$6.1\pm 4.6$ \\
$10^3{\cal B}_{\Sigma^{+} \rho^{0}}$
&$6.1\pm4.6$  \\
$10^3{\cal B}_{\Xi^{0} K^{*+}}$
&$8.7 \pm 2.7$ \\
$10^3{\cal B}_{\Lambda^{0} \rho^{+}}$
&$7.4\pm 3.4$\\
\hline

\end{tabular}
%=====================
\begin{tabular}{|c|c|}
\hline
$\Xi_c^0$&our results
\\
\hline
$10^4{\cal B}_{\Sigma^{+} \rho^{-}}$
&$5.6 \pm 1.8$ \\
$10^4{\cal B}_{\Sigma^{-} \rho^{+}}$
&$5.3 \pm 0.7$ \\
$10^5{\cal B}_{\Sigma^{0} \rho^{0}}$
&$8.2 \pm 6.7$ \\
$10^4{\cal B}_{\Sigma^{0} \omega}$
&$1.0 \pm 0.8$ \\
$10^4{\cal B}_{\Sigma^{0} \phi}$
&$2.4 \pm1.1$ \\
$10^4{\cal B}_{\Xi^{-} K^{*+}}$
&$3.9 \pm 0.5$ \\
$10^4{\cal B}_{\Xi^{0} K^{*0}}$
&$6.3 \pm 2.0$ \\
$10^4{\cal B}_{p K^{*-}}$
&$3.0\pm 2.2$ \\
$10^4{\cal B}_{n \bar{K}^{*0}}$
&$4.5 \pm 3.4$ \\
$10^4{\cal B}_{\Lambda^{0} \rho^{0}}$
&$9.2 \pm 2.2$ \\
$10^4{\cal B}_{\Lambda^{0} \omega}$
&$0.1^{+2.5}_{-0.1}$ \\
\hline\hline
$\Xi_c^+$&our results
\\\hline
$10^3{\cal B}_{\Sigma^{0} \rho^{+}}$
&$1.9 \pm 0.6$ \\
$10^3{\cal B}_{\Sigma^{+} \rho^{0}}$
&$1.9 \pm 0.6$ \\
$10^4{\cal B}_{\Sigma^{+} \omega}$
&$8.2\pm5.9$ \\
$10^3{\cal B}_{\Sigma^{+} \phi}$
&$1.9\pm0.9$ \\
$10^4{\cal B}_{\Xi^{0} K^{*+}}$
&$9.6 \pm7.9$ \\
$10^3{\cal B}_{p \bar{K}^{*0}}$
&$7.8 \pm 2.2$ \\
$10^3{\cal B}_{\Lambda^{0} \rho^{+}}$
&$7.1 \pm1.7$ \\
\hline\hline
$\Lambda_c^+$&our results
\\\hline
$10^4{\cal B}_{p \rho^{0}}$
&$3.5 \pm 2.9$ \\
$10^4{\cal B}_{n \rho^{+}}$
&$7.0 \pm 5.8$ \\
$10^3{\cal B}_{\Sigma^{0} K^{*+}}$
&$1.2\pm 0.3$\\
$10^3{\cal B}_{\Lambda^{0} K^{*+}}$
&$2.0 \pm 0.5$ \\
\hline
\end{tabular}
%=====================
\begin{tabular}{|c|c|}
\hline
$\Xi_c^0$&our results
\\
\hline
$10^5{\cal B}_{p \rho^{-}}$
&$3.6 \pm 1.1$ \\
$10^5{\cal B}_{\Sigma^{-} K^{*+}}$
&$2.5 \pm 0.3$ \\
$10^5{\cal B}_{\Sigma^{0} K^{*0}}$
&$1.3 \pm 0.2$ \\
$10^5{\cal B}_{n \rho^{0}}$
&$1.8 \pm 0.6$ \\
$10^5{\cal B}_{n \omega}$
&$9.9\pm1.6$ \\
$10^5{\cal B}_{n \phi}$
&$3.7\pm1.8$\\
$10^4{\cal B}_{\Lambda^{0} K^{*0}}$
&$8.1\pm 7.2$\\
&\\&\\&\\&\\
\hline\hline
$\Xi_c^+$&our results
\\\hline

$10^5{\cal B}_{\Sigma^{0} K^{*+}}$
&$5.0\pm 0.7$\\
$10^5{\cal B}_{\Sigma^{+} K^{*0}}$
&$9.9 \pm 1.3$ \\
$10^5{\cal B}_{p \rho^{0}}$
&$7.1 \pm 2.2$ \\
$10^4{\cal B}_{p \omega}$
&$3.9\pm0.6$ \\
$10^4{\cal B}_{p \phi}$
&$1.5\pm0.7$ \\
$10^4{\cal B}_{n \rho^{+}}$
&$1.4\pm 0.4$ \\
$10^5{\cal B}_{\Lambda^{0} K^{*+}}$
&$3.2 \pm 2.9$ \\
\hline\hline
$\Lambda_c^+$&our results
\\\hline
$10^4{\cal B}_{p K^{*0}}$
&$1.6 \pm 0.5$ \\
$10^4{\cal B}_{n K^{*+}}$
&$1.6 \pm 0.5$\\
&\\
&\\
\hline
\end{tabular}
}
\end{table}
%==================================
%==================================
%\newpage
\section{Discussions and Conclusions}
With $\chi^2/n.d.f\simeq 2$ to present a reasonable fit,
the approach based the $SU(3)_f$ symmetry is demonstrated
to be reliable for ${\bf B}_c\to {\bf B}_n V$.
Besides, our prediction 
\begin{eqnarray}\label{1st_pre}
{\cal B}(\Lambda^{+}_{c}\to \Sigma^{+}\rho^{0},\Lambda^0 \rho^+)
&=&(0.61\pm 0.46,0.74\pm 0.34)\%\,,
\end{eqnarray}
agrees with the experimental upper bounds of $(1.7,6)\%$,
respectively~\cite{pdg}. 
We also find
\begin{eqnarray}
&&{\cal B}(\Lambda^+_c \to 
\Xi^0 K^{*+},\Sigma^0 K^{*+},\Lambda^0 K^{*+})
=(8.7 \pm 2.7,1.2\pm 0.3,2.0\pm 0.5)\times 10^{-3}\,,
\end{eqnarray}
to be compatible with the pseudo-scalar counterparts.
According to Table~\ref{tab1}, we obtain 
\begin{eqnarray}
&&T(\Lambda_c^+\to\Sigma^{0} \rho^{+})
+T(\Lambda_c^+\to\Sigma^{+} \rho^{0})=0\,,\nonumber\\
&&T(\Lambda_c^+\to\Sigma^{+} K^{*0} )
-\sqrt 2 T(\Lambda_c^+\to\Sigma^{0} K^{*+})=-2\bar a_5 s_c\,,\nonumber\\
&&T(\Lambda_c^+\to n\rho^+)
-\sqrt 2 T(\Lambda_c^+\to  p \rho^{0} )=
-(\frac{\bar a_4+\bar a_6}{2})s_c\,,\nonumber\\
&&T(\Lambda_c^+\to nK^{*+})
+T(\Lambda_c^+\to p {K}^{*0})=-2(\bar a_4+\bar a_5)s_c^2\,.
%&&T(\Lambda_c^+\to \Sigma^{+} \phi)=\bar a_4-2\bar h+\bar h'\,.
\end{eqnarray}
By ignoring the parameters in $H(\overline{15})$,
we obtain
\begin{eqnarray}
&&{\cal B}(\Lambda_c^+\to\Sigma^{0} \rho^{+},\Sigma^{+} \rho^{0})
=(6.1\pm4.6)\times 10^{-3}\,,\nonumber\\
&&{\cal B}(\Lambda_c^+\to  p \rho^{0} )=\frac{1}{2}{\cal B}(\Lambda_c^+\to n\rho^+)
=(3.5 \pm 2.9)\times 10^{-4}\,,\nonumber\\
&&{\cal B}(\Lambda_c^+\to nK^{*+}, p {K}^{*0}) 
=(1.6 \pm 0.5)\times 10^{-4}\,,
\end{eqnarray}
which respect the isospin symmetry.
%${\cal B}(\Lambda_c^+\to\Sigma^{+} \phi)$ is shown to
%only receive the $h$ term. 
We also get
\begin{eqnarray}\label{re2}
&&
\frac{1}{\sqrt 2}T(\Lambda_c^+\to p \bar {K}^{*0})-
\frac{1}{s_c}T(\Lambda_c^+\to  p \rho^{0})
=T(\Lambda_c^+\to\Sigma^{0} \rho^{+})\,,\nonumber\\
&&
\frac{1}{\sqrt 2}T(\Lambda_c^+\to p \bar {K}^{*0})+
\frac{1}{s_c}T(\Lambda_c^+\to  p \rho^{0})
=\sqrt 3T(\Lambda_c^+\to\Lambda^{0} \rho^{+})\,,
\end{eqnarray}
which lead to
\begin{eqnarray}\label{prho0}
{\cal B}(\Lambda_c^+\to p \rho^0)&\simeq &\frac{s_c^2}{2}
[3.6{\cal B}(\Lambda_c^+\to \Lambda^0 \rho^+)+1.3{\cal B}(\Lambda_c^+\to \Sigma^0 \rho^+)
-1.1{\cal B}(\Lambda_c^+\to p \bar K^{*0})]\,,
\end{eqnarray}
where the pre-factors (3.6,1.3,1.1) have
taken into account the differences for $|\vec{p}_{cm}|$ in Eq.~(\ref{p_space}).
It is interesting to note that the $\Lambda_c^+\to p \pi^0$ decay
has a similar relation to that in Eq.~(\ref{prho0}),
where $(\rho,\bar K^{*0})$ are replaced by $(\pi,\bar K^{0})$.
However, the relation for $\Lambda_c^+\to p \pi^0$ causes 
${\cal B}(\Lambda_c^+\to p \pi^0)\simeq 5\times 10^{-4}$,
disapproved by the data~\cite{pdg}. This indicates that, 
even though the ignoring of $H(\overline{15})$ is viable,
the possible interferences between $H(6)$ and $H(\overline{15})$ 
might give sizeable contributions to some decay modes~\cite{Geng:2018rse}.
In this work, since the fit still accommodates the data, 
it is not clear 
which of the $\Lambda_c^+\to {\bf B}_n V$ decays receives
sizeable interferences between $H(6)$ and $H(\overline{15})$.
Like the ${\cal B}(\Lambda_c^+\to p \pi^0)$ case, 
the precise measurement of ${\cal B}(\Lambda_c^+\to p \rho^0)$
can test the ignoring of $H(\overline{15})$. 
For the $\Xi_c^+$ decays, we obtain 
\begin{eqnarray}
&&
{\cal B}(\Xi^+_c \to \Sigma^+\bar K^{*0},\Xi^{0} \rho^{+})
=(10.1\pm 2.9,9.9 \pm2.9)\times 10^{-2}\,,\nonumber\\
&&
{\cal B}(\Xi^+_c \to p \bar K^{*0},\Sigma^+ \phi)
=(7.8 \pm 2.2,1.9\pm0.9)\times 10^{-3}\,.
\end{eqnarray}
With $f_{\tau_{{\bf B}_c}}\equiv \tau_{\Xi_c^+}/\tau_{\Lambda_c^+}\simeq 2.2$,
${\cal B}(\Xi^+_c \to \Sigma^+\bar K^{*0},\Xi^{0} \rho^{+})
\simeq (2-4)f_{\tau_{{\bf B}_c}}{\cal B}(\Lambda_c^+ \to p \bar K^{*0})$
is found to be in accordance with $|\bar a_3|^2\simeq (2-4)|\bar a_1|^2$, 
which can be tested by more accurate measurements.

By means of ${\cal B}({\bf B}_c\to {\bf B}_n V,V\to MM')=
{\cal B}({\bf B}_c\to {\bf B}_n V){\cal B}(V\to MM')$, 
the resonant contribution to the total ${\cal B}({\bf B}_c\to {\bf B}_n MM')$
can be investigated, where $MM'$ from the vector meson decay
are in the P-wave state. On the other hand,
the theoretical study of ${\bf B}_c\to{\bf B}_n MM'$
needs $MM'$ to be mainly in the S-wave state~\cite{Geng:2018upx}.
Using ${\cal B}(\rho^{0(+)}\to\pi^+\pi^{-(0)})\simeq 100\%$~\cite{pdg} and 
the predictions for ${\cal B}(\Lambda_c^+\to \Sigma\rho,\Lambda^0\rho^+)$,
we obtain 
\begin{eqnarray}
{\cal B}(\Lambda_c^+\to \Sigma^+ \rho^0,\rho^0\to\pi^+\pi^-)
&=&(6.1\pm 4.6)\times 10^{-3}\,,\nonumber\\
{\cal B}(\Lambda_c^+\to \Sigma^0 \rho^+,\rho^+\to\pi^+\pi^0)
&=&(6.1\pm 4.6)\times 10^{-3}\,,\nonumber\\
{\cal B}(\Lambda_c^+\to \Lambda^0\rho^+,\rho^+\to\pi^+\pi^0)
&=&(7.4\pm 3.4)\times 10^{-3}\,,
\end{eqnarray}
which are within the total branching ratios of 
$(4.42\pm 0.28,2.2\pm 0.8,7.0\pm 0.4)\times 10^{-2}$~\cite{pdg},
respectively, showing that the P-wave contributions from
$V\to MM'$ are indeed minor to these decays.
By putting ${\cal B}( \Xi^+_c \to p\phi)=(1.5\pm 0.7)\times 10^{-4}$
into the measured ratio of 
${\cal B}( \Xi^+_c \to p\phi)/{\cal B}(\Xi^+_c \to pK^- \pi^+)
=(19.8\pm 0.7\pm 0.9\pm 0.2)\times 10^{-3}$~\cite{Aaij:2019kss},
we obtain ${\cal B}(\Xi^+_c \to pK^- \pi^+)=(0.8\pm 0.4)\%$,
which is a little smaller than the predicted value 
of $(1.7\pm 0.5)\%$~\cite{Wang:2019dls}.

In sum, within the framework of the $SU(3)_f$ symmetry,
we have studied the ${\bf B}_c\to{\bf B}_n V$ decays.
We have predicted 
${\cal B}(\Lambda^{+}_{c}\to \Sigma^{+}\rho^{0},\Lambda^0 \rho^+)
=(0.61\pm 0.46,0.74\pm 0.34)\%$, 
in agreement with the experimental upper bounds of $(1.7,6)\%$,
respectively. 
It has also been shown that 
${\cal B}(\Lambda^+_c \to 
\Xi^0 K^{*+},\Sigma^0 K^{*+},\Lambda^0 K^{*+})
=(8.7 \pm 2.7,1.2\pm 0.3,2.0\pm 0.5)\times 10^{-3}$.
For the $\Xi_c^+$ decays, we have obtained 
${\cal B}(\Xi^+_c \to \Sigma^+\bar K^{*0},\Xi^{0} \rho^{+})
=(10.1\pm 2.9,9.9 \pm2.9)\times 10^{-2}$,
${\cal B}(\Xi^+_c \to p \bar K^{*0},\Sigma^+ \phi)
=(7.8 \pm 2.2,1.9\pm0.9)\times 10^{-3}$ and
${\cal B}( \Xi^+_c \to p\phi)=(1.5\pm 0.7)\times 10^{-4}$.
The predicted ${\cal B}({\bf B}_c\to{\bf B}_n V)$ can be compared to
the future measurements by BESIII, BELLEII and LHCb.

\section*{ACKNOWLEDGMENTS}
This work was supported by
National Science Foundation of China (11675030).


\begin{thebibliography}{99}

\bibitem{pdg}
M.~Tanabashi {\it et al.} [Particle Data Group],
%``Review of Particle Physics,''
Phys.\ Rev.\ D {\bf 98}, 030001 (2018).

\bibitem{Ablikim:2018czr}
M.~Ablikim {\it et al.} [BESIII Collaboration],
%``Evidence for the decays of $\Lambda^+_{c}\to\Sigma^+\eta$ and $\Sigma^+\eta^\prime$,''
arXiv:1811.08028 [hep-ex].

\bibitem{Li:2018qak}
Y.B.~Li {\it et al.} [Belle Collaboration],
%``First measurements of absolute branching fractions of $\Xi_c^0$ at Belle,''
arXiv:1811.09738 [hep-ex].

\bibitem{Geng:2018upx} 
C.Q.~Geng, Y.K.~Hsiao, C.W.~Liu and T.H.~Tsai,
%``Three-body charmed baryon Decays with SU(3) flavor symmetry,''
arXiv:1810.01079 [hep-ph].

\bibitem{Aaij:2019kss}
R.~Aaij {\it et al.} [LHCb Collaboration],
%``Observation of the doubly Cabibbo-suppressed decay $\Xi_{c}^{+}\to p\phi$,''
arXiv:1901.06222 [hep-ex].

% factorization=====
\bibitem{ali} A. Ali, G. Kramer and C.D. Lu, Phys. Rev.  D{\bf 58}, 094009 (1998).


\bibitem{Geng:2006jt} 
C.Q.~Geng, Y.K.~Hsiao and J.N.~Ng,
%``Direct CP violation in B+- ---> p anti-p K(*)+-,''
Phys.\ Rev.\ Lett.\  {\bf 98}, 011801 (2007).

\bibitem{Hsiao:2014mua} 
Y.K.~Hsiao and C.Q.~Geng,
%``Direct CP violation in $\Lambda_b$ decays,''
Phys.\ Rev.\ D {\bf 91}, 116007 (2015).
%=========


\bibitem{Zhao:2018mov}
H.J.~Zhao, Y.K.~Hsiao and Y.~Yao,
%``A diagrammatic analysis of two-body charmed baryon decays with flavor symmetry,''
arXiv:1811.07265 [hep-ph].

%SU3_B:====
\bibitem{He:2000ys} 
X.G.~He, Y.K.~Hsiao, J.Q.~Shi, Y.L.~Wu and Y.F.~Zhou,
%``The CP violating phase $\gamma$ from global fit of rare charmless hadronic $B$ decays,''
Phys.\ Rev.\ D {\bf 64}, 034002 (2001).

\bibitem{Fu:2003fy} 
H.K.~Fu, X.G.~He and Y.K.~Hsiao,
%``B ---> eta(eta-prime) K(pi) in the standard model with flavor symmetry,''
Phys.\ Rev.\ D {\bf 69}, 074002 (2004).

\bibitem{Hsiao:2015iiu} 
Y.K.~Hsiao, C.F.~Chang and X.G.~He,
%``A global $SU(3)/U(3)$ flavor symmetry analysis for $B\to PP$ with $\eta-\eta'$ Mixing,''
Phys.\ Rev.\ D {\bf 93}, 114002 (2016).

%SU3_Lb:====
\bibitem{He:2015fwa} 
X.G.~He and G.N.~Li,
%``Predictive $CP$ violating relations for charmless two-body decays of 
%beauty baryons $\Xi^{-,\;0}_b$ and $\Lambda_b^0$ with flavor $SU(3)$ symmetry,''
Phys.\ Lett.\ B {\bf 750}, 82 (2015).

\bibitem{He:2015fsa} 
M.~He, X.G.~He and G.N.~Li,
%``CP-Violating Polarization Asymmetry in Charmless Two-Body Decays of Beauty Baryons,''
Phys.\ Rev.\ D {\bf 92}, 036010 (2015).
%===========

% SU3 in Lc:====
\bibitem{Savage:1989qr} 
M.J.~Savage and R.P.~Springer,
%``SU(3) Predictions for Charmed Baryon Decays,''
Phys.\ Rev.\ D {\bf 42}, 1527 (1990).

\bibitem{Savage:1991wu} 
M.J.~Savage,
%``SU(3) violations in the nonleptonic decay of charmed hadrons,''
Phys.\ Lett.\ B {\bf 257}, 414 (1991).

\bibitem{h_term}
G.~Altarelli, N.~Cabibbo and L.~Maiani,
%``Weak Nonleptonic Decays of Charmed Hadrons,''
Phys.\ Lett.\  {\bf 57B}, 277 (1975).

%\cite{}
\bibitem{He:2018joe} 
X.G.~He, Y.J.~Shi and W.~Wang,
%``Unification of Flavor SU(3) Analyses of Heavy Hadron Weak Decays,''
arXiv:1811.03480 [hep-ph].

\bibitem{Lu:2016ogy} 
C.D.~Lu, W.~Wang and F.S.~Yu,
%``Test flavor SU(3) symmetry in exclusive $\Lambda_c$ decays,''
Phys.\ Rev.\ D {\bf 93}, 056008 (2016).

\bibitem{Wang:2017azm} 
W.~Wang, Z.P.~Xing and J.~Xu,
%``Weak Decays of Doubly Heavy Baryons: SU(3) Analysis,''
Eur.\ Phys.\ J.\ C {\bf 77}, 800 (2017). %arXiv:1707.06570 [hep-ph].

\bibitem{Wang:2017gxe}
D.~Wang, P.F.~Guo, W.H.~Long and F.S.~Yu,
%``$K_S^0-K_L^0$ Asymmetries and $CP$ Violation in Charmed Baryon Decays into Neutral Kaons,''
JHEP {\bf 1803}, 066 (2018). %arXiv:1709.09873 [hep-ph].

\bibitem{Geng:2017esc} 
C.Q.~Geng, Y.K.~Hsiao, Y.H.~Lin and L.L. Liu,
%``Non-leptonic two-body weak decays of $\Lambda_c(2286)$,''
Phys.\ Lett.\ B {\bf 776}, 265 (2018). %arXiv:1708.02460 [hep-ph].

%\cite{}
\bibitem{Geng:2018plk} 
C.Q.~Geng, Y.K.~Hsiao, C.W.~Liu and T.H.~Tsai,
%``Antitriplet charmed baryon decays with SU(3) flavor symmetry,''
Phys.\ Rev.\ D {\bf 97}, 073006 (2018). %[arXiv:1801.03276 [hep-ph]].

\bibitem{Geng:2018bow} 
C.Q.~Geng, Y.K.~Hsiao, C.W.~Liu and T.H.~Tsai,
 %``SU(3) symmetry breaking in charmed baryon decays,''
Eur.\ Phys.\ J.\ C {\bf 78}, 593 (2018). %[arXiv:1804.01666 [hep-ph]].

%\cite{}
\bibitem{Geng:2018rse} 
C.Q.~Geng, C.W.~Liu and T.H.~Tsai,
%``Singly Cabibbo suppressed decays of $\Lambda_{c}^+$ with SU(3) flavor symmetry,''
Phys.\ Lett.\ B {\bf 790}, 225 (2019). %[arXiv:1812.08508 [hep-ph]].
%=============================

%II. Formalism
\bibitem{Buras:1998raa} %``Weak Hamiltonian, CP violation and rare decays,''
A.J.~Buras, hep-ph/9806471.

\bibitem{Fajfer:2002gp} 
S.~Fajfer, P.~Singer and J.~Zupan,
%``The Radiative leptonic decays D0 ---> e+ e- gamma, mu+ mu- gamma 
%in the standard model and beyond,''
Eur.\ Phys.\ J.\ C {\bf 27}, 201 (2003).

\bibitem{Li:2012cfa} 
H.n.~Li, C.D.~Lu and F.S.~Yu,
%``Branching ratios and direct CP asymmetries in $D\to PP$ decays,''
Phys.\ Rev.\ D {\bf 86}, 036012 (2012).

%III. Numerical analysis=================
%\cite{}
\bibitem{Link:2003cd}
J.M.~Link {\it et al.} [FOCUS Collaboration],
%``Measurements of Xi(c)+ branching ratios,''
Phys.\ Lett.\ B {\bf 571}, 139 (2003). %[hep-ex/0305038].

\bibitem{Aaij:2017nsd}
R.~Aaij {\it et al.} [LHCb Collaboration],
%``Search for the rare decay $\Lambda_{c}^{+} \to p\mu^+\mu^-$,''
Phys.\ Rev.\ D {\bf 97}, 091101 (2018).


%\cite{Ablikim:2018bir}
\bibitem{Ablikim:2018bir}
M.~Ablikim {\it et al.} [BESIII Collaboration],
%``Measurements of absolute branching fractions for $\Lambda^+_c\to\Xi^0K^+$ and $\Xi(1530)^0K^+$,''
Phys.\ Lett.\ B {\bf 783}, 200 (2018). %[arXiv:1803.04299 [hep-ex]].

%\cite{Kubota:1993pw}
\bibitem{Kubota:1993pw}
Y.~Kubota {\it et al.} [CLEO Collaboration],
%``Measurement of exclusive Lambda(c) decays with a Sigma+ in the final state,''
Phys.\ Rev.\ Lett.\  {\bf 71}, 3255 (1993).

%\cite{Aaij:2017nsd}
\bibitem{Aaij:2017nsd}
R.~Aaij {\it et al.} [LHCb Collaboration],
%``Search for the rare decay $\Lambda_{c}^{+} \to p\mu^+\mu^-$,''
Phys.\ Rev.\ D {\bf 97}, 091101 (2018).

\bibitem{Wang:2019dls} 
D.~Wang,
%``Sum rules for $CP$ asymmetries of charmed baryon decays in the flavor $SU(3)$ limit,''
arXiv:1901.01776 [hep-ph].

\end{thebibliography}
\end{document}